\documentstyle[twoside,epsf]{article}

\catcode`\@=11
\long\def\@makefntext#1{
\protect\noindent \hbox to 3.2pt {\hskip-.9pt  
$^{{\eightrm\@thefnmark}}$\hfil}#1\hfill}		

\def\@makefnmark{\hbox to 0pt{$^{\@thefnmark}$\hss}}	
	
\def\ps@myheadings{\let\@mkboth\@gobbletwo
\def\@oddhead{\hbox{}
\rightmark\hfil\eightrm\thepage}   
\def\@oddfoot{}\def\@evenhead{\eightrm\thepage\hfil
\leftmark\hbox{}}\def\@evenfoot{}
\def\sectionmark##1{}\def\subsectionmark##1{}}

\oddsidemargin=\evensidemargin
\newcounter{sectionc}\newcounter{subsectionc}\newcounter{subsubsectionc}
\renewcommand{\section}[1] {\vspace{12pt}\addtocounter{sectionc}{1} 
\setcounter{subsectionc}{0}\setcounter{subsubsectionc}{0}\noindent 
	{\tenbf\thesectionc. #1}\par\vspace{5pt}}
\renewcommand{\subsection}[1] {\vspace{12pt}\addtocounter{subsectionc}{1} 
	\setcounter{subsubsectionc}{0}\noindent 
	{\bf\thesectionc.\thesubsectionc. {\kern1pt \bfit #1}}\par\vspace{5pt}}
\renewcommand{\subsubsection}[1] {\vspace{12pt}\addtocounter{subsubsectionc}{1}
	\noindent{\tenrm\thesectionc.\thesubsectionc.\thesubsubsectionc.
	{\kern1pt \tenit #1}}\par\vspace{5pt}}
\newcommand{\nonumsection}[1] {\vspace{12pt}\noindent{\tenbf #1}
	\par\vspace{5pt}}

\newcommand{\textlineskip}{\baselineskip=13pt}
\newcommand{\smalllineskip}{\baselineskip=10pt}

\def\eightcirc{
\begin{picture}(0,0)
\put(4.4,1.8){\circle{6.5}}
\end{picture}}
\def\eightcopyright{\eightcirc\kern2.7pt\hbox{\eightrm c}} 

\newcommand{\copyrightheading}[1]
	{\vspace*{-2.5cm}\smalllineskip{\flushleft
      {\footnotesize Mod. Phys. Lett. A 16 (2001) 2359-2362  #1}\\
       {\footnotesize $\eightcopyright$\, World Scientific Publishing Company
        }\\
	 }}

\def\abstracts#1#2#3{{
	\centering{\begin{minipage}{4.5in}\baselineskip=10pt\footnotesize
	\parindent=0pt #1\par 
	\parindent=15pt #2\par
	\parindent=15pt #3
	\end{minipage}}\par}} 

\renewenvironment{thebibliography}[1]
	{\frenchspacing
	 \ninerm\baselineskip=11pt
	 \begin{list}{\arabic{enumi}.}
        {\usecounter{enumi}\setlength{\parsep}{0pt}     
	 \setlength{\leftmargin 12.7pt}{\rightmargin 0pt} 
         \setlength{\itemsep}{0pt} \settowidth
	{\labelwidth}{#1.}\sloppy}}{\end{list}}

\newcounter{itemlistc}
\newcounter{romanlistc}
\newcounter{alphlistc}
\newcounter{arabiclistc}

\def\@citex[#1]#2{\if@filesw\immediate\write\@auxout
	{\string\citation{#2}}\fi
\def\@citea{}\@cite{\@for\@citeb:=#2\do
	{\@citea\def\@citea{,}\@ifundefined
	{b@\@citeb}{{\bf ?}\@warning
	{Citation `\@citeb' on page \thepage \space undefined}}
	{\csname b@\@citeb\endcsname}}}{#1}}

\newif\if@cghi
\def\cite{\@cghitrue\@ifnextchar [{\@tempswatrue
	\@citex}{\@tempswafalse\@citex[]}}
\def\citelow{\@cghifalse\@ifnextchar [{\@tempswatrue
	\@citex}{\@tempswafalse\@citex[]}}
\def\@cite#1#2{{$\null^{#1}$\if@tempswa\typeout
	{IJCGA warning: optional citation argument 
	ignored: `#2'} \fi}}

\def\@refcitex[#1]#2{\if@filesw\immediate\write\@auxout
	{\string\citation{#2}}\fi
\def\@citea{}\@refcite{\@for\@citeb:=#2\do
	{\@citea\def\@citea{, }\@ifundefined
	{b@\@citeb}{{\bf ?}\@warning
	{Citation `\@citeb' on page \thepage \space undefined}}
	\hbox{\csname b@\@citeb\endcsname}}}{#1}}

\def\@refcite#1#2{{#1\if@tempswa\typeout
        {IJCGA warning: optional citation argument
	ignored: `#2'} \fi}}

\def\refcite{\@ifnextchar[{\@tempswatrue
	\@refcitex}{\@tempswafalse\@refcitex[]}}

\def\pmb#1{\setbox0=\hbox{#1}
	\kern-.025em\copy0\kern-\wd0
	\kern.05em\copy0\kern-\wd0
	\kern-.025em\raise.0433em\box0}

\def\fnt#1#2{\footnotetext{\kern-.3em
	{$^{\mbox{\scriptsize #1}}$}{#2}}}

\def\runninghead#1#2{\pagestyle{myheadings}
\markboth{{\protect\footnotesize\it{\quad #1}}\hfill}
{\hfill{\protect\footnotesize\it{#2\quad}}}}
\font\tenrm=cmr10
\font\tenit=cmti10 
\font\tenbf=cmbx10
\font\bfit=cmbxti10 at 10pt
\font\ninerm=cmr9

\font\eightrm=cmr8

\def\qed{\hbox{${\vcenter{\vbox{			
   \hrule height 0.4pt\hbox{\vrule width 0.4pt height 6pt
   \kern5pt\vrule width 0.4pt}\hrule height 0.4pt}}}$}}

\begin{document}

\newpage

\runninghead{Haret C. Rosu, Jose Torres} 
{Grassmann Hubble superfield}

\normalsize\textlineskip
\thispagestyle{empty}
\setcounter{page}{1}

\copyrightheading{}    

\vspace*{0.88truein}

\bigskip
\centerline{\bf A GRASSMANN REPRESENTATION OF THE HUBBLE PARAMETER}  
\vspace*{0.035truein}
\vspace*{0.37truein}
\vspace*{10pt}
\centerline{\footnotesize HARET C. ROSU and JOSE TORRES}
\vspace*{0.015truein}
\centerline{\footnotesize  Instituto de F\'{\i}sica,
Universidad de Guanajuato, Apdo Postal E-143, 37150 Le\'on, Gto, Mexico}
\vspace*{0.225truein}

\vspace*{0.21truein}
\abstracts{The Riccati equation for the Hubble parameter $H$ of
barotropic FRW cosmologies in conformal
time for $\kappa \neq 0$ spatial geometries and in comoving time for the 
$\kappa =0$ geometry, respectively, is generalized to odd Grassmannian time 
parameters. We obtain 
a system of simple differential equations for the four supercomponents 
(two of even type and two of odd type) of the Hubble superfield 
function ${\cal H}$ that is explicitly solved. 
The second even Hubble component does not have an evolution governed by general
relativity although there are effects of the latter upon it. }{}{}


\textlineskip                  
\vspace*{12pt}                 

\vspace*{1pt}\textlineskip	
\vspace*{-0.5pt}
\noindent


\noindent





{\bf 1} - Recently, Faraoni,\cite{Far} showed that the equations 
describing barotropic FRW cosmologies can be combined in a simple 
Riccati equation leading in a much easier way 
to the textbook solutions for the FRW scale factors.
Faraoni obtained the following cosmological Riccati equation
$$
 \frac{dH}{d\eta}=-cH^2-\kappa c~,
\eqno(1a)
$$
for the log derivative of the FRW scale factor, the famous Hubble
parameter $H(\eta)=\frac{da/d\eta}{a}$.
The independent variable is the conformal time $\eta$, 
$c=\frac{3}{2}\gamma -1$ was assumed constant by Faraoni, 
and $\kappa=0,\pm1$ is the curvature index of the flat, closed, open
FRW universe, respectively. The adiabatic index $\gamma$ is defined through 
the barotropic equation of state $P=(\gamma -1)\rho$, where $P$ and $\rho$
are, respectively, the pressure and energy density of the cosmological fluid 
under consideration. It is a constant quantity in the approach of Faraoni.
It is a simple matter to write
the first Friedmann dynamical equation $\frac{\ddot{a}}{a}=-\frac{4\pi G}{3}
(\rho +P)$ in the form $\frac{\ddot{a}}{a}=-\frac{4\pi G\rho}{3}c$ for the 
barotropic case. Thus, the common lore is that
a negative $c$ ($\gamma < \frac{2}{3}$)
implies an accelerating universe as
claimed in the astrophysics of supernovae.  

We also notice that from the mathematical point of view, Faraoni`s FRW 
Riccati equation being of constant coefficients is directly integrable. 
The solutions are 
$$
H^+=-\tan c\eta ~, \qquad H^-={\rm coth}(c\eta)~,
\eqno(1b)
$$ 
 for the closed and open FRW universes, respectively.

In the following we perform a very simple generalization of the 
cosmological Riccati equation
($1a$) by considering the Hubble parameter as a function not only of time 
but also  of the so-called odd ``time" parameters, $\theta _1$ and 
$\bar{\theta}_1$
(where $\bar{\theta}_1$ is the complex conjugate of $\theta _1$).\cite{tk}
In this way, the Hubble parameter becomes a superfield in the terminology of 
superanalysis.\cite{ber} This is only a toy generalization since we do not
relate it to any sophisticated formalism, such as supergravity. However, 
we obtain some simple analytic formulas for all the components
of the Hubble superfield. We mention that similar types of 
supersymmetric extensions (affecting in addition the operatorial structure)
are of much interest in the theory of integrable 
models that appear naturally in the study of strings in the matrix 
model approach.\cite{das}

\bigskip

{\bf 2} - 
We shall write the Riccati equation for the superfield ${\cal H}$ in the 
following form
$$
\frac{d{\bf {\cal H}}}{d\eta}\equiv \frac{i}{2}
\{ D_{\theta _1},D_{\bar{\theta}_1}\}{\bf {\cal H}}
=\alpha {\bf{\cal H}}^2 + \beta~,
\eqno(2)
$$
where
$$
{\bf {\cal H}}(\eta , \theta _1, \bar{\theta}_1)=H_0(\eta) +i\theta _1 
\bar{H}_1(\eta)+
i\bar{\theta}_1H_1(\eta) +\theta _1 \bar{\theta}_1H_{11}(\eta) ~,
\eqno(3)
$$
$\alpha =-c$, $\beta =-\kappa c$, and 
$$
D_{\theta _1}=\frac{d}{d\theta _1}+i\bar{\theta}_1\frac{d}{d\eta}~, \qquad 
D_{\bar{\theta}_1}=-\frac{d}{d\bar{\theta} _1}-i\theta _1\frac{d}{d\eta}~.  
\eqno(4)
$$
The superfield ${\bf {\cal H}}$ satisfies the reality condition 
${\bf {\cal H}}^{\dagger} ={\bf {\cal H}}$ and is
assumed to be dimensionless implying the following $\eta$ dimensions
$$
[\eta]=1~,\qquad [\theta _{1}]=1/2~, \qquad [H_0]=0~,\qquad 
[H_1]=-1/2~, \qquad [H_{11}]=-1~.
$$
After simple (super)calculations we get the following system of equations
$$
\frac{dH _0}{d\eta}=\alpha H_{0}^{2}+\beta,
\eqno(5a)
$$
$$
\frac{d\bar{h} _1}{d\eta}=2\alpha H_{0}\bar{h} _1,
\eqno(5b)
$$
$$
\frac{dh _1}{d\eta}=2\alpha H_{0}h _1,
\eqno(5c)
$$
$$
\hskip 1.6cm\frac{dH _{11}}{d\eta}=2\alpha ( H_{0}H_{11}+\bar{h} _1h _1
\bar{\epsilon}\epsilon),
\eqno(5d)
$$
where $\bar{h}_1\bar{\epsilon}=\bar{H}_1$,  $h_1\epsilon =H_1$, and 
$\epsilon$ is a Grassmann parameter.

 
One can see that equation (5a) 
is identical to Faraoni's Riccati equation.
Therefore, the solutions are
$$
H_{0}^{+}=-{\rm tan}(c\eta), \qquad H_{0}^{-}={\rm coth}(c\eta),
\eqno(6)
$$
For the other components, we get simple first order differential equations.
The equations (5b) and (5c) are identical since we consider only the 
possibility of real superpartners. The solutions are easily obtained 
$$
h_1^+=\frac{1}{\cos ^2(c\eta)}~,\qquad h_1^-=\frac{1}{{\rm sinh} ^2(c\eta)}~, 
\eqno(7)
$$
for the closed and open spatial geometries, respectively.

Finally, the solution of the equation for $H_{11}$ can be obtained by the 
method of the variation of constants leading to
$$
H_{11}^{\pm}=e^{2\alpha\int _{\eta _0}^{\eta} H_0 d\eta}\Bigg[A_{11}^{\pm}+
\bar{\epsilon}\epsilon
\int _{\eta _0}^{\eta}2h_1^2e^{-2\alpha\int ^{\eta} H_0d\eta}d\eta\Bigg]~,
\eqno(8)
$$
where the parameters $A_{11}^{\pm}$ are related to the initial conditions.
We thus obtain
$$
H_{11}^{+}=\cos ^{-2}(c\eta)\left( A_{11}^{+}-
2\bar{\epsilon}\epsilon{\rm tan}(c\eta)
\right)
\eqno(9)
$$
and
$$
H_{11}^{-}={\rm sinh} ^{-2}(c\eta)\left( A_{11}^{-}+
2\bar{\epsilon}\epsilon {\rm coth}(c\eta)
\right)~.
\eqno(10)
$$

\bigskip

{\bf 3} - The case $\kappa =0$ is special in the sense that $\alpha =
-(c+1)$, $\beta =0$ 
and the independent variable is the comoving time $t$.\cite{0} Therefore,
the odd time parameters refer to the comoving time in this case.

\bigskip

 (i) For $c+1\neq 0$ ($\alpha \neq 0, \, \beta =0$) we get
$$
H_0^0=\frac{1}{c+1}\frac{1}{t},\qquad 
h_1^0=\bar{h}^{0}_{1}=\frac{1}{t^2}~,\qquad
H_{11}^{0}=A_{11}^{0}t^{-2}
+2\bar{\epsilon}\epsilon (c+1)
t^{-3}~.
\eqno(11)
$$

\bigskip 

 (ii) The case $c+1=0$ ($\alpha = 0, \, \beta =0$) leads to
$$
H_0=C_0, \qquad h_1=\bar{h}_{1}=C_1, \qquad H_{11}
=C_{11}~.
\eqno(12)
$$


\bigskip



\bigskip

{\bf 4} - Through a simple generalization of the barotropic FRW Riccati 
evolution equation 
to odd Grassmann time parameters ($\theta _1$, $\bar{\theta}_1$) 
a Grassmann
(superfield) representation of the Hubble parameter is given in this work. The 
superfield components are explicitly
obtained under the simple assumption of a real Hubble superfield.
We notice that the evolution equation of the second even component
$H_{11}$ is a 
linear first order differential equation not a nonlinear Riccati equation as 
required by
Friedmann equations. Thus, we are out of the realm of general relativity
although there is an induced effect of the latter as seen from the presence of
$H_0$ in equation (5d).

\nonumsection{Acknowledgements}

\noindent
The authors thank Prof. V. Tkach and Dr. J.J. Rosales for discussions.


\end{document}